\documentclass[
reprint,
superscriptaddress,
amsmath,
amssymb,
aps,
longbibliography,
prb,
showpacs,
floatfix
]{revtex4-2}
\usepackage{graphicx} 
\usepackage[dvipsnames]{xcolor}
\usepackage[bookmarks=false,linkcolor=Orange,urlcolor=MidnightBlue,colorlinks,citecolor=Maroon]{hyperref}
\usepackage{physics}
\usepackage{comment}
\usepackage[normalem]{ulem}

\begin{document}
\title{Anomalous phonon magnetic moments}
\author{Swati Chaudhary}
\altaffiliation{These authors contributed equally to this work.}
\email{swatichaudhary@issp.u-tokyo.ac.jp}
\affiliation{The Institute for Solid State Physics, The University of Tokyo, Kashiwa, Chiba 277-8581, Japan}

\author{Carl P. Romao}
\altaffiliation{These authors contributed equally to this work.}
\email{carl.romao@cvut.cz}
\affiliation{Department of Materials, ETH Zürich, Wolfgang-Pauli-Str. 27, 8093 Zürich, Switzerland}
\affiliation{Department of Materials, Faculty of Nuclear Sciences and Physical Engineering, Czech Technical University in Prague, Trojanova 13, Prague 120 00, Czech Republic}

\author{Dominik M. Juraschek}
\affiliation{Department of Applied Physics and Science Education, Eindhoven University of Technology, 5612 AP Eindhoven, Netherlands}


\begin{abstract}
Circularly polarized phonons conventionally carry an angular momentum and a magnetic moment arising from circular motions of the atoms. Here, we present three anomalous cases that lead to phonon magnetic moments, which cannot be described in the conventional framework: \textit{rotationless axial phonons}, which exhibit magnetic responses despite only carrying pseudo angular momentum, \textit{divergent gyromagnetic ratios of phonons}, in which a magnetic moment is produced despite vanishing angular momentum, and {\textit{anisotropic gyromagnetic ratios of phonons}, which make the phonon angular momentum and magnetic moment noncollinear.} Our results shed light on the origin and nature of phonon magnetism and suggest the existence of phononomagnetic hidden order.
\end{abstract}

\maketitle


\section{Introduction}

Circularly polarized phonons carry angular momentum that can be exchanged with other particles, such as photons or electrons, on fundamental timescales~\cite{Zhu2018,Tauchert2022,Davies2024}. These lattice vibrations, appearing in the form of axial and chiral phonons \cite{ueda2023chiral,ishito2023truly, uedaChiral2025,Juraschek2025}, produce magnetic moments and hence represent a fundamental degree of freedom in crystalline solids in addition to electronic angular momentum \cite{Juraschek2019,Ren2021,Zhang2023gate,chaudhary2024giant}. The phonon magnetic moment, $\mathbf{m}^{ph}$, and angular momentum, $\textbf{l}^{ph}$, have so-far been considered to be collinear and proportional to each other by a phonon gyromagnetic ratio, $\mathbf{m}^{ph}=\gamma^{ph} \mathbf{l}^{ph}$~\cite{Juraschek2019}.

The origin of angular momentum of a (quasi)particle is not always directly linked to circular motion, as is commonly exemplified by the electron spin. An alternative definition of the angular momentum can be formulated from the transformation of a wavefunction under rotation~\cite{Zhang2015PRL,Streib2021}. 
For example, the variation of the phase of the electromagnetic field of light with an azimuthal angle gives rise to a twisted wavefront and therefore orbital angular momentum even for linear polarization~\cite{AllenPRA1992}. 
The same principle also applies to elastic deformations and lattice vibrations  with azimuthal phase variation \cite{Gao2023,Garanin2015,Nakane2018}, as well as to twisted magnon beams~\cite{jia2019twisted}. On the atomic length scale, phonon angular momentum can be obtained by acting the $n$-fold rotation operator on the phonon displacement vector \cite{Zhang2015PRL,Wang2024}. This form is known as phonon pseudo angular momentum (PAM) and stems from a phase difference between the motion of different atoms belonging to the same unit cell (spin PAM) or different unit cells (orbital PAM)~\cite{Zhang2015PRL}. The prototypical example of phonons carrying PAM is found in the $K$/$K'$ valleys of hexagonal lattices~\cite{Zhang2015PRL,Zhu2018,Chen2019,chen2019chiralkagome,yao2024conversion,ptok2021chiral}. There, the angular momentum from circular motion and that from relative phase correlations between different atoms are intertwined. The literature to date has not considered the case of phonon eigenmodes where angular momentum arises purely from phase correlations at atomic length scales without any circular motion of atoms. 

\begin{figure}[b]
    \centering
    \includegraphics[width=\linewidth]{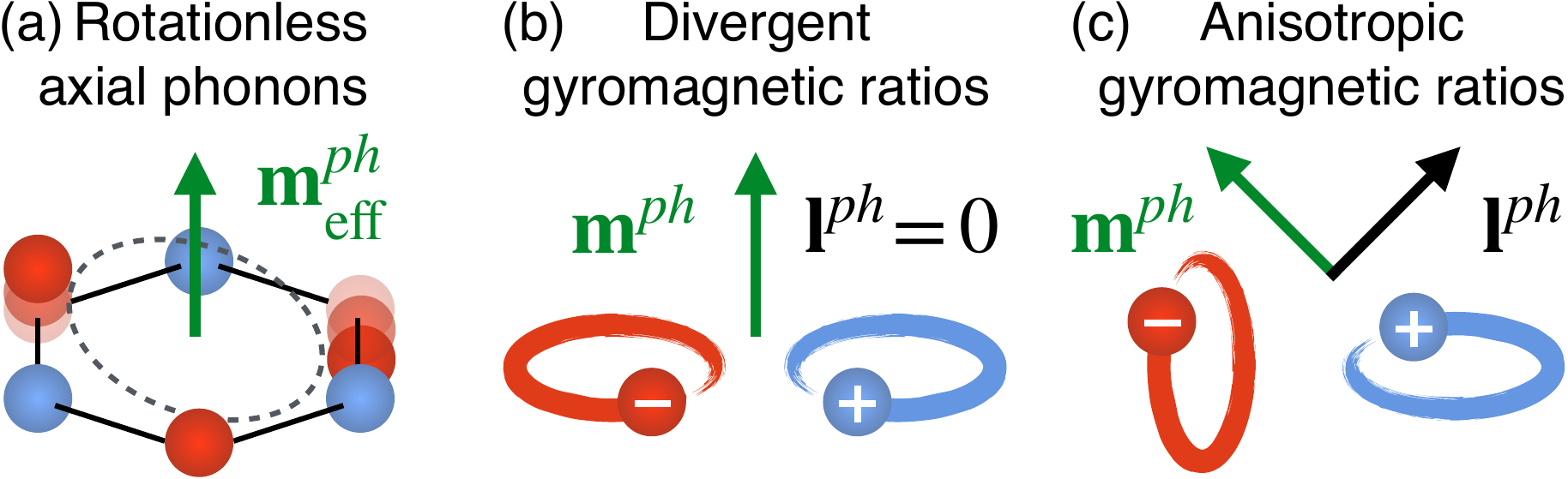}
    \caption{{Anomalous phonon magnetic moments. (a) Rotationless axial phonons can carry effective magnetic moments while containing only linear atomic motion. (b) Divergent gyromagnetic ratios produce finite magnetic moments despite vanishing angular momentum. (c) Anisotropic gyromagnetic ratios produce noncollinear phonon angular momentum and magnetic moments.}}
    \label{fig:overview}
\end{figure}

In this work, we demonstrate three cases that defy the conventional picture of the phonon magnetic moment: {firstly, \textit{rotationless axial phonons}, which possess only pseudo, but no real angular momentum, and which can generate effective phonon magnetic moments; secondly, phonons with divergent gyromagnetic ratios, in which a finite magnetic moment is generated despite vanishing angular momentum; and thirdly, anisotropic gyromagnetic ratios that cause the phonon angular momentum and magnetic moment to be noncollinear. A schematic overview is shown in Fig.~\ref{fig:overview}.} Because our analysis will involve both chiral and achiral phonons carrying angular momentum, we will in the following use the neutral term ``axial phonons'' to indicate they can be represented by the axial vector $\mathbf{l}$ \cite{Juraschek2025}.


\section{Phonon angular momentum and magnetic moments}

We begin by reviewing the formalism of phonon angular momentum. 
Real phonon angular momentum arises from circular motion of atoms around their equilibrium positions in a crystal and, for a specific phonon mode $\nu$ at a wavevector $\mathbf{q}$, can be written as \cite{zhang:2014}
\begin{equation}
    \mathbf{l}^{ph}_{\nu \mathbf{q}} = \hbar \sum_{\alpha} \mathbf{l}^{ph}_{\nu\mathbf{q}\alpha} =   \sum_{\alpha}\mathbf{u}_{\nu\mathbf{q}\alpha}\times \dot{\mathbf{u}}{_{\nu\mathbf{q}\alpha}},
\label{eq:phononangmom}
\end{equation}
where $\mathbf{u}_{\nu\mathbf{q}\alpha}$ is the phonon displacement vector for atom $\alpha$. In ionic crystals, these phonons further carry a magnetic moment produced by a circular charge current of the ions that can be written as \cite{ueda2023chiral, Juraschek2019,Geilhufe2021,Zabalo2022}
\begin{align}
    \mathbf{m}^{ph}_{\nu\mathbf{q}} = \sum_\alpha \mathbf{m}^{ph}_{\nu\mathbf{q}\alpha} =  \hbar \sum_\alpha \frac{Z^*_{\alpha}}{2 M_\alpha} \mathbf{l}^{ph}_{\nu\mathbf{q}\alpha},
\label{eq:phononmagmom}
\end{align}
where $Z^*_{\alpha}$ is the Born effective charge tensor, $M_\alpha$ the atomic mass, and $Z^*_{\alpha}/(2M_\alpha)$ the gyromagnetic ratio of atom $\alpha$. This magnetic moment arises from circular charge currents and produces a Maxwellian magnetic field from the dynamical multiferroic effect \cite{juraschek2:2017,merlin2025magnetophononics}. Its magnitude is typically on the order of the nuclear magneton, $\mu_n$ \cite{Juraschek2019,ueda2023chiral, romao2023chiral,Zhang2023gate, uedaChiral2025}, but can be enhanced by up to four orders of magnitude through Maxwellian or non-Maxwellian contributions from electron- and spin-phonon coupling~\cite{Ren2021,chaudhary2024giant,Merlin2024,Mustafa2025,Xiao-Wei2023,hernandez2023observation,geilhufe2023electron,geilhufe2022inertia,lujan2024spin}.

Phonons can further carry pseudo angular momentum (PAM) arising from phase differences between atomic motions under $n$-fold rotational operations~\cite{Zhang2015PRL},
\begin{equation}
   C_n(z)\mathbf{u}_{\nu\mathbf{q}} e^{i\mathbf{R}_{\alpha l} \cdot\mathbf{q}}=e^{-i\frac{2\pi}{n}\hat{z}\cdot\mathbf{l}^p_{\nu\mathbf{q}}}\mathbf{u}_{\nu\mathbf{q}} e^{i\mathbf{R}_{\alpha l} \cdot\mathbf{q}},
    \label{Eq:rotationOAM}
\end{equation}
where $\mathbf{l}^p_{\nu\mathbf{q}}$ is the PAM along the rotation axis $\hat{z}$ passing through a $n-$fold symmetric point in the unit cell and can take values of $l^p_{\nu\mathbf{q}}=0,...,(n-1)$, and $\mathbf{R}_{\alpha l}$ is the position vector for atom $\alpha$ in unit cell $l$. The phase factor can arise from a rotation of the displacement vector $\mathbf{u}_{\nu\mathbf{q}}$ directly (intracell, spin PAM), or from a rotation of the nonlocal part, $e^{i\mathbf{R}_j \cdot\mathbf{q}}$, (intercell, orbital PAM), for which $l^p_{\nu\mathbf{q}} = l^{p,s}_{\nu\mathbf{q}} + l^{p,o}_{\nu\mathbf{q}}$. This quantized PAM arises at points in the Brillouin zone that respect $n$-fold rotational symmetry.

Real phonon angular momentum has been associated with the generation of magnetic fields~\cite{juraschek2:2017,Geilhufe2021,Shabala2024,luo2023large,nova2017effective,Juraschek2022_giantphonomag,basini2024terahertz,Davies2024}, whereas PAM is important for selection rules in light- and electron-phonon scattering~\cite{Zhang2015PRL,Zhu2018,Chen2019_entangle,ueda2023chiral,ishito2023truly,zhang2025chirality}. In these and other studies, the investigated phonons involve circular motions of atoms and exhibit collinear phonon angular momentum and magnetic moments. 
PAM in turn arises from the action of the rotational operator $C_n$ and makes no assumptions about the circularity of the atomic motion. Hence, also purely linear motions of the atoms along the rotation axis should lead to PAM as long as there is a phase difference between them.


\section{Rotationless axial phonons}

\begin{figure}
    \centering
    \includegraphics[scale=0.34]{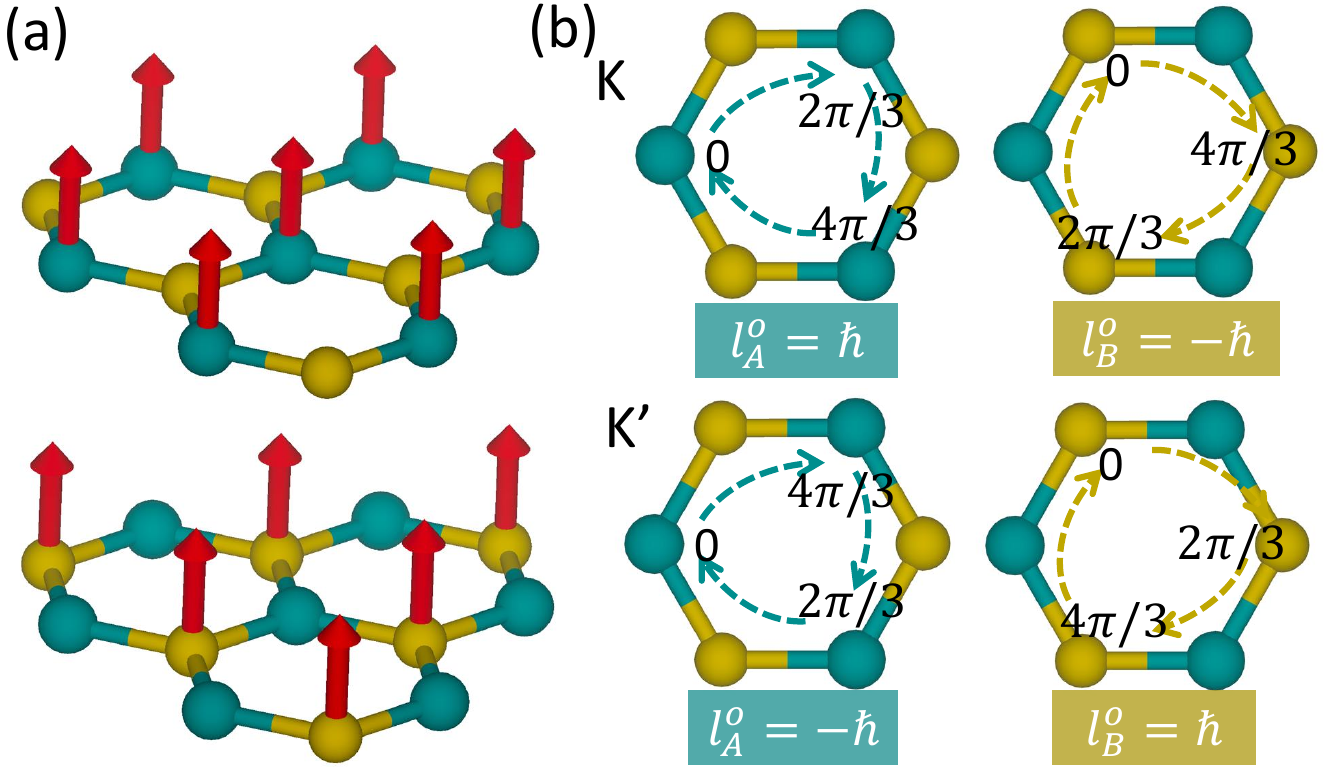}
    \caption{Out-of-plane atomic displacements in monolayer h-BN. (a) Atomic displacement associated with the transverse acoustic modes, which correspond to the out-of-plane motion from phonons at $K/K'$ valleys with frequencies of 9.2 THz (top) and 17.9 THz (bottom). While all eigenvectors point out of the plane, their relative motion is phase delayed as shown in (b), resembling the motion of an Euler disk. (b) $K$ and $K'$ valley phonons have an opposite phase difference, resulting in an orbital PAM of $\pm 1$. Boron atoms are shown in green and nitrogen atoms are shown in yellow.}
    \label{figh-BN}
\end{figure}

{We next introduce rotationless axial phonons in hexagonal materials for the simple examples of honeycomb and kagome lattices. We then derive the magnetic response for hexagonal cerium trichloride (CeCl$_3$), and show that they produce effective magnetic moments purely from PAM, despite vanishing $\mathbf{l}^{ph}_{\nu\mathbf{q}}$ and $\mathbf{m}^{ph}_{\nu\mathbf{q}}$. As we will be considering phonon modes of specific branches and wavevectors, we will drop the indices $\nu$ and $\mathbf{q}$.
}

\subsection{Rotationless axial phonons in honeycomb lattices}
     The three-fold rotation around the center of the hexagon leads to orbital PAM that is opposite in sign for different sublattices, $A$ and $B$, $l^{p,o}(A)=-l^{p,o}(B)=\pm1$ for all phonon branches at $\mathbf{q}=K/K'$. Additionally, $K/K'$-valley phonons with in-plane circular motion also carry a spin PAM of $l^{p,s}=\pm 1$, resulting in a total PAM of $l^p=l^{p,s}(A)+l^{p,o}(A)=l^{p,s}(B)+l^{p,o}(B)$. The three-fold rotation symmetry enforces a total PAM of $l^p=\pm1,0$ modulo 3, which constrains the in-plane valley phonons to have only one sublattice moving or to exhibit opposite circular motion for the $A$ and $B$ sublattices. On the other hand, phonon modes with linear out-of-plane motion cannot carry any spin PAM, implying a total PAM of $l^{p}=l^{p,o}=\pm1$. Given that $l^{p,o}(A)=-l^{p,o}(B)=\pm1$, any nondegenerate valley phonon with linear out-of-plane motion must have only one sublattice moving and the total PAM about the three-fold axis passing through the center of the hexagon or through the stationary sublattice would be $\pm1$. This 
     requires broken inversion symmetry and is realized in hexagonal boron nitride (h-BN). 
We investigate out-of-plane nondegenerate transverse acoustic $K/K'$ phonons in monolayer h-BN in Fig.~\ref{figh-BN} (a), calculated for h-BN from first principles \cite{gonze2020abinit, perdew1996generalized, grimme2010consistent, sdata}. (For computational details, see Appendix~A.) These phonon modes involve only the motion of one sublattice and pick up a phase of $\pm2\pi/3$ after three-fold rotation around the center of honeycomb, as shown in Fig.~\ref{figh-BN} (b), resulting in an orbital PAM of $\pm1$ arising from the intercell phase difference.

\subsection{Rotationless axial phonons in kagome lattices}
\begin{figure}
    \centering
    \includegraphics[scale=0.55]{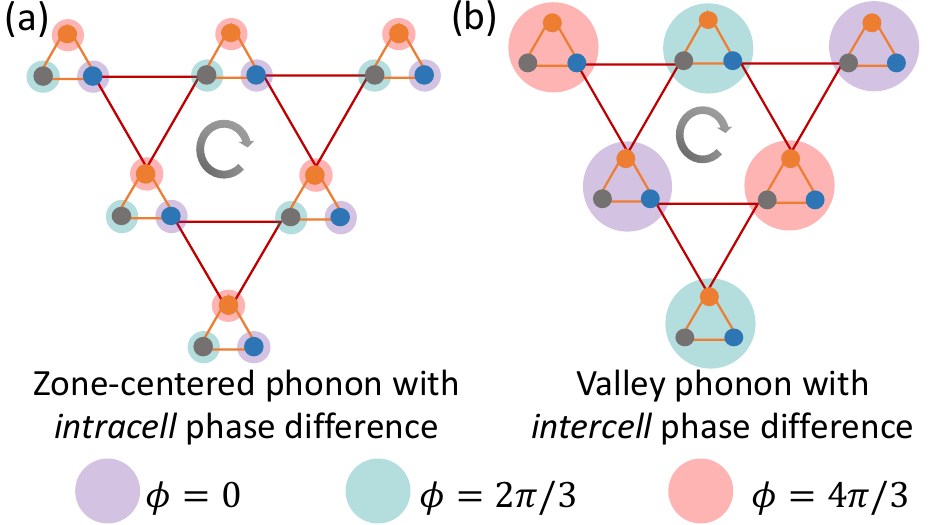}
    \caption{Spin and orbital PAM in a kagome lattice. Phonons with out-of-plane motion in a kagome lattice can carry PAM due to intracell and intercell phase differences. 
    A three-fold rotation of the phonon mode around a $C_3$ symmetric point in the lattice results in a 
    phase of $\pm2\pi/3$. (a) Zone-center phonon, producing a $2\pi/3$ phase difference between the three atoms on a given lattice site, leading to a spin PAM of $1$ with the same sign for all sublattices under a three-fold rotation. (b) Valley phonons, with a $2\pi/3$ phase difference between the three lattice sites from neighboring unit cells, resulting in an orbital PAM under a three-fold rotation.}
    \label{figkagome}
\end{figure}

In kagome lattices, each site contains three atoms related by three-fold rotation about an axis passing through the center of hexagon or through the center of the triangle in the kagome structure. 
This implies equal orbital PAM for all sublattices and allows spin PAM arising from the intracell phase difference for phonon modes with out-of-plane linear motion.  
Such phonon modes can carry both spin PAM and orbital PAM, as illustrated in Fig.~\ref{figkagome}. 
For $K/K'$-valley phonons, a three-fold rotation around the center of the hexagon results in an overall phase of $2\pi/3$ due to the phase difference between different unit cells as shown by different colors in Fig.~\ref{figkagome}(b).

Similarly, for the $\Gamma$-point phonon, the three-fold rotation around the center of hexagon or triangle in the kagome unit results in the same phase of $2\pi/3$ when different sublattices are moving out-of-plane with a phase difference of $2\pi/3$, as shown in Fig.~\ref{figkagome}(a). From group theoretical analysis, we expect that such a scenario arises for the $E_u$ modes in the kagome magnet Co$_3$Sn$_2$S$_2$ that involves the motion of Co ions along the $z$ axis of the crystal (see Appendix~B for details). Previous studies have already suggested that $E_g$ and $E_u$ modes carrying real angular momentum in this material split in an applied magnetic field and hence possess a magnetic moment \cite{che2024magnetic,yang2024inherent}. Another example are the $E_{2u}$ modes in FeGe that should exhibit similar properties. These phonons can become axial when time-reversal symmetry is broken either due to an applied magnetic field or magnetic ordering. 

\begin{figure*}[t]
    \centering
    \includegraphics[scale=0.47]{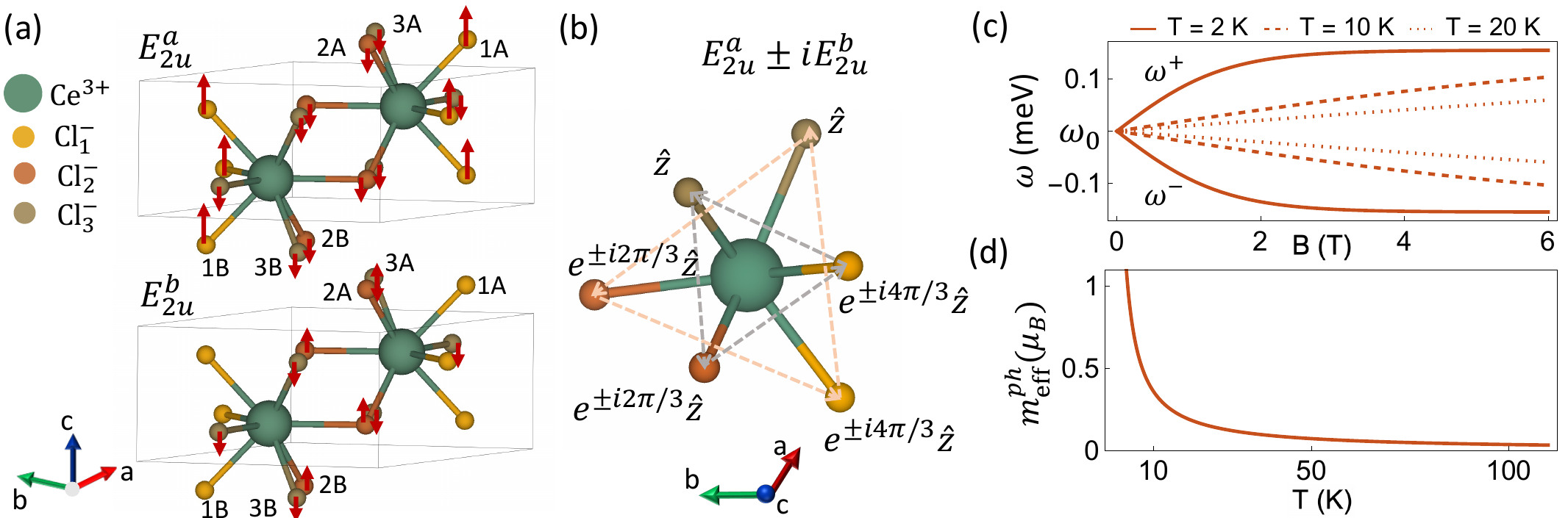}
    \caption{Axial $E_{2u}$ mode and its magnetic response. (a) Displacement associated with the two orthogonal components of the $E_{2u}$ mode. (b) A circular superposition of the two components results in a relative phase difference between displacements of different atoms within the same unit cell, leading to a phonon spin pseudo angular momentum. (c) Phonon Zeeman splitting, $\Delta\omega_{ph}$ of the axial $E_{2u}$ mode in the presence of an external magnetic field. (d) Temperature dependence of the effective phonon magnetic moment, $m_\mathrm{eff}^{ph}$.}
    \label{figCeCl3Eg}
\end{figure*}

\subsection{Magnetic response of rotationless axial phonons in cerium trichloride}
For rotationless zone-center phonons, spin PAM can arise from an intracell phase difference, as we have shown for kagome lattices above. We now show that these phonons can carry effective magnetic moments {and respond to applied magnetic fields}, despite carrying no real magnetic moments according to Eq.~\eqref{eq:phononmagmom}. We use the example of the rare-earth $4f$-paramagnet CeCl$_3$, for which giant phono-magnetic and magneto-phononic effects have been discovered due to its strong orbit-lattice coupling \cite{schaack:1976,schaack:1977,Juraschek2022_giantphonomag,luo2023large,chaudhary2024giant}. The material crystallizes in the hexagonal space group $P6_3/m$ 
and exhibits one $E_{1g}$ and one $E_{2u}$ symmetric phonon mode at the center of the Brillouin zone that induce purely out-of-plane atomic motions, as illustrated in Fig.~\ref{figCeCl3Eg}. (For details on CeCl$_3$, see Appendix~D.)

The phonon displacement vector of the $E_{2u}$ mode at 20.5~meV can be obtained from group theory~\cite{bcs:sam},
\begin{align}
     \mathbf{u}_{E_{2u}}^a&=\frac{Q_a}{2\sqrt{6}}\left(0,0, 2 \hat{z},-\hat{z},-\hat{z},2 \hat{z},-\hat{z},-\hat{z}\right),\label{E1gmodead}\\
     \mathbf{u}_{E_{2u}}^b&=\frac{Q_b}{2\sqrt{2}}\left(0,0,0,\hat{z},-\hat{z},0,\hat{z},-\hat{z}\right),\label{E1gmodebd}
\end{align}
with basis (Ce$_A^{3+}$,Ce$_B^{3+}$,Cl$_{1A}^{-}$,Cl$_{2A}^{-}$,Cl$_{3A}^{-}$,Cl$_{1B}^{-}$,Cl$_{2B}^{-}$,Cl$_{3B}^{-}$), and where $Q_{a/b}$ are the normal mode coordinates (amplitudes) of the two orthogonal components $a$ and $b$ in units of \AA$\sqrt{\mathrm{amu}}$, where amu is the atomic mass unit. The circular superposition of two components of the $E_{2u}$ mode, $\textbf{Q}^{\pm}= \mathbf{u}_{E_{2u}}^a\pm i \mathbf{u}_{E_{2u}}^b$, results in axial phonon given by 
\begin{equation}
\textbf{Q}^{\pm}=\frac{Q}{\sqrt{6}}\hat{z}\left(0,0,1,e^{\pm i\frac{2\pi}{3}},e^{\pm i\frac{4\pi}{3}},1,e^{\pm i\frac{2\pi}{3}},e^{\pm i\frac{4\pi}{3}}\right),
\end{equation}
which indicates that the three Cl$^{-}$ ions in a given $xy$ plane around the Ce$^{3+}$ ions are moving in $z$ direction but with a relative phase difference of $\pm 2\pi/3$ as shown in Fig.~\ref{figCeCl3Eg}(b). This leads to spin PAM of $l^{p,s}=\pm 1$ as evident by the three-fold rotation on the phonon mode:
\begin{equation}
    C_3(z)\textbf{Q}^{\pm} = e^{ -i\frac{2\pi}{3}\, l^{p,s}}\textbf{Q}^{\pm}.
\end{equation}

It is known that in the presence of a magnetic field, phonons with real angular momentum can exhibit the Zeeman effect, which in turn can be used to define an effective phonon magnetic moment. We now calculate the splitting of these axial phonons in the presence of an applied magnetic field along the direction of three-fold rotation axis. We use the formalism developed in Ref.~\cite{chaudhary2024giant} to compute the phonon frequency splitting arising from orbit-lattice coupling which depends on the polarization state of phonon eigenmodes. In this mechanism, phonons hybridize with orbital excitations on the magnetic ion that carry finite angular momentum, resulting in a frequency splitting that depends on the phonon angular momentum, see Appendixes~C and D for details.  Here, we find the phonon eigenmodes, $\textbf{Q}^{\pm}$ which carry opposite PAM, exhibit a frequency splitting that depends on the magnetic field and temperature, as shown in Fig.~\ref{figCeCl3Eg}(c). It is linear for small $B$ and saturates to a value of $0.3$~meV for increasing magnetic fields. In contrast to Eq.~\eqref{eq:phononmagmom}, we can use this PAM-dependent phonon frequency splitting to define an effective {gyromagnetic ratio} and phonon magnetic moment as
\begin{equation}
    m^{ph}_\mathrm{eff}=\gamma^{ph}_\mathrm{eff}l^{ph}=\frac{1}{2}\,\frac{\partial \Delta\omega_{ph}}{\partial B}\bigg|_{B=0},
    \label{eq:effectivephononmagmom}
\end{equation}
where $\Delta\omega_{ph}$ denotes the energy splitting (in meV) of axial phonons with opposite PAM due to the applied magnetic field. Using Eq.~\eqref{eq:effectivephononmagmom}, we find an effective phonon magnetic moment of the order of $1~\mu_B$ at a temperature of $T=2$~K, which decreases monotonically with temperature as shown in Fig.~\ref{figCeCl3Eg}(d). {While the effective phonon magnetic moment characterizes the phononic response to an applied magnetic field, it also produces a real magnetic moment arising from spin polarization in the paramagnetic material that is given by $m^{ph}_\mathrm{real}=\chi_m m^{ph}_\mathrm{eff}$, where $\chi_m$ is the magnetic susceptibility. Because $\chi_m\sim10^{-4}$ \cite{luo2023large}, the real magnetic moment is on the order of the nuclear magneton.} The phonon Zeeman effect has not yet been measured for infrared-active or silent modes in CeCl$_3$, however the rotationless Raman-active $E_{1g}$ mode was shown to exhibit a strong Zeeman splitting in early experiments~\cite{schaack:1977}. {Our results demonstrate that circular atomic motion and therefore real phonon angular momentum is neither a necessary requirement for electron spin to phonon angular momentum coupling, nor for magnetic responses of axial phonons.}


Now that we have shown that effective phonon magnetic moments can arise solely from PAM, 
we will consider two more anomalous behaviors of phonon modes, in which the magnetic moment is not directly proportional to the real angular momentum.


\section{Divergent gyromagnetic ratios of phonons}

\begin{figure}
\centering
    \includegraphics[width=0.48\textwidth]{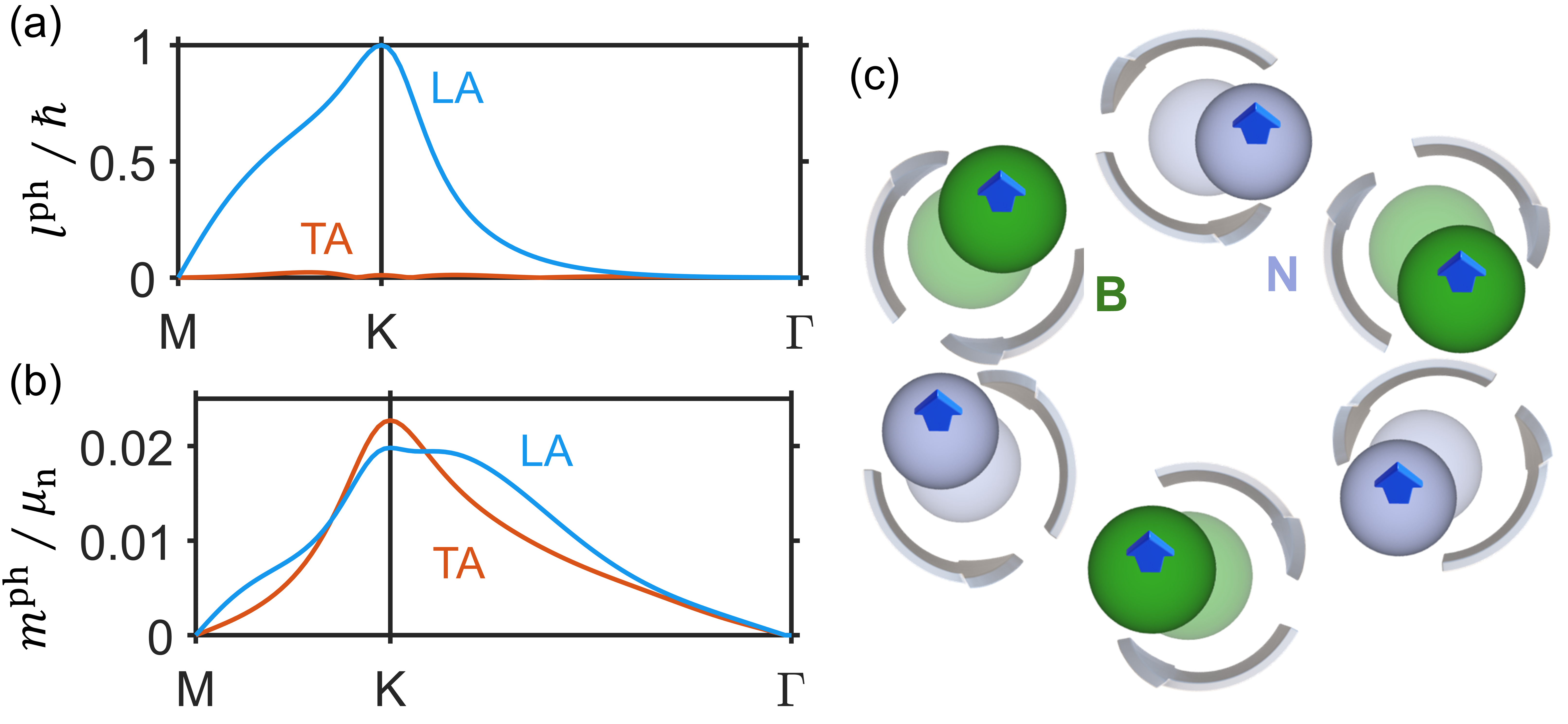}
    \caption{(a) Phonon angular momentum, $l^{ph}$, and (b) phonon magnetic moment, $m^{ph}$, (in units of the nuclear magneton, $\mu_\mathrm{n}$) of the longitudinal acoustic (LA) and fast transverse acoustic (TA) bands of monolayer h-BN, shown as a function of wavevector along the $\mathrm{M} \ (\frac{1}{2} \ 0 \ 0)-\mathrm{K} \ (\frac{1}{3} \ \frac{1}{3} \ 0) - \mathrm{\Gamma} \ (0 \ 0 \ 0)$ trajectory in reciprocal space. (c) Atomic displacements in the TA band at the $\mathrm{K}$ point, where B and N atoms revolve in opposite directions around their equilibrium positions (transparent spheres). This leads to a near-vanishing net angular momentum, counter-aligned for the two sublattices, but a substantial net magnetic moment (blue arrows), co-aligned for the two sublattices due to their opposite effective charges.}
    \label{figBN}
\end{figure}

Axial phonons exhibit both circular and, as we have shown, linear motions of the atoms \cite{romao2023chiral}. If ionic sublattices revolve in opposite directions, phonons without net angular momentum can still produce phonon magnetism. For example, in a crystal with one cation ($+$) and one anion ($-$) per unit cell, the effective charges are equal in magnitude and opposite in sign, $Z^*_{+}=-Z^*_{-}$. For a phonon mode with $\mathbf{l}^{ph}_{+} = -\mathbf{l}^{ph}_{-}$, the phonon angular momentum in Eq.~\eqref{eq:phononangmom} vanishes, whereas the phonon  magnetic moment in Eq.~\eqref{eq:phononmagmom} remains nonzero. Accordingly, the phonon gyromagnetic ratio, defined as $\gamma^{ph} = \frac{|\mathbf{m}^{ph}|}{|\mathbf{l}^{ph}|}$, diverges. This ideal case is approximated in monolayer h-BN. Axial phonons have previously been studied in h-BN \cite{Rostami2022} and in heterostructures containing h-BN \cite{li2021, gao2018nondegenerate}, where they show a phonon band structure similar to the noncentrosymmetric transition metal dichalcogenides \cite{Zhang2015PRL}. Here, we demonstrate the existence of near-divergent gyromagnetic ratios in Fig.~\ref{figBN} using the calculated phononic properties of h-BN \cite{gonze2020abinit, perdew1996generalized, grimme2010consistent, sdata}.

As shown in Fig.~\ref{figBN}a, the longitudinal acoustic (LA) phonon branch is fully circularly polarized at $\mathbf{q}=K$, corresponding to circular motions of the B sublattice, while the fast transverse acoustic (TA) branch has nearly vanishing phonon angular momentum. However, the TA branch still possesses a substantial phonon magnetic moment (Fig.~\ref{figBN}b) due to the clockwise motions of the B cations combined with the counterclockwise motions of the N anions (Fig. \ref{figBN}c), leading to a near divergence of the gyromagnetic ratio. 


\section{Anisotropic gyromagnetic ratios of phonons}

Divergent gyromagnetic ratios of phonons can be expanded to the more general phenomenon of anisotropic gyromagnetic ratios of phonons. This phenomenon occurs when the atomic phonon angular momentum vectors, $\mathbf{l}_\alpha^{ph}$, of the different ionic sublattices are not aligned. We show here that this misalignment occurs at arbitrary $\mathbf{q}$ even in a highly symmetric crystal with only two independent atomic positions. The misalignment leads to a noncollinear local arrangement of the magnetic moments arising from circular atomic motions, $\mathbf{m}^{ph}_{\alpha}$, and a difference in spatial orientation between the phonon angular momentum and phonon magnetic moment vectors. Accordingly, both quantities are connected by a tensorial phonon gyromagnetic ratio, 
$\mathbf{m}^{ph} = \underline{\underline{\gamma}}^{ph} \mathbf{l}^{ph}$, where $\mathbf{m}^{ph}\nparallel\mathbf{l}^{ph}$.

\begin{figure}
\centering
    \includegraphics[width=0.48\textwidth]{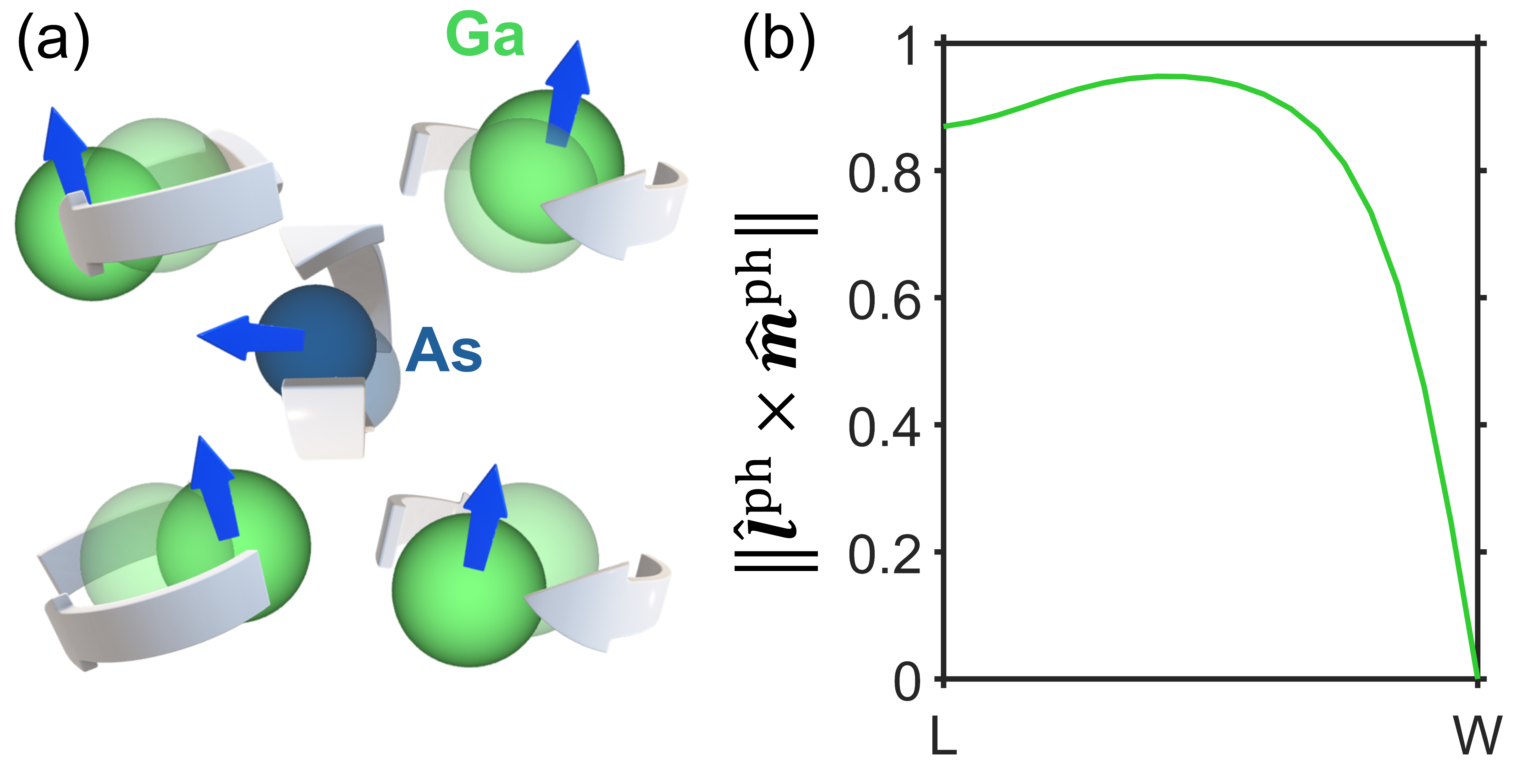}
    \caption{(a) The noncollinear nature of the atomic moments (blue arrows) in GaAs, arising from the orbital motions of the atoms (spheres) around their average positions (transparent spheres), is shown schematically for a phonon halfway between  $L \ (\frac{1}{2} \ \frac{1}{2} \ \frac{1}{2})$ and $W \ (\frac{1}{2} \ \frac{1}{4} \ \frac{3}{4})$. This noncollinearity gives rise to a difference between the spatial orientation of the phonon angular momentum, $\mathbf{l}^{ph}$, and magnetic moment vectors, $\mathbf{m}^{ph}$. (b) Magnitude of the cross product of the unit vectors, $\hat{\mathbf{l}}^{ph}\times\hat{\mathbf{m}}^{ph}$, for the highest energy optical branch along the $L - W$ trajectory, showing the effect of the anisotropic phonon gyromagnetic ratio.}
    \label{figGaAs}
\end{figure}

We demonstrate anisotropic gyromagnetic ratios of phonons for the example of noncentrosymmetric gallium arsenide (GaAs) in Fig.~\ref{figGaAs}, using phonon data computed from first principles \cite{jain2013commentary, osti_1200591}, for details see Appendix~A. The optical phonons between the $L$ and $W$ points involve noncoplanar circular motions of the atoms. The magnetic moments of the Ga ions are primarily aligned along the $z$ axis of the crystal, while those of the As ions are oriented in the $xy$ plane (Fig. \ref{figGaAs}a). In combination with the differing gyromagnetic ratios of the two ions, a phonon magnetic moment emerges that is nearly orthogonal to the phonon angular momentum along most of the $L-W$ direction (Fig. \ref{figGaAs}b).

\section{Discussion}

The rotationless axial phonons presented here challenge the conventional understanding that circular atomic motion is required to induce spin PAM and phonon magnetic moments and shows that both can be achieved with only linear atomic motion. The large effective magnetic moment of the rotationless $E_{2u}$ mode in CeCl$_3$ suggests that coherent excitation with ultrashort mid-infrared pulses can induce strong effective magnetic fields, akin to those from conventional circularly polarized phonons~\cite{Juraschek2020,Juraschek2022_giantphonomag,luo2023large}. These phonons may also couple to intrinsic orders with atomic scale phase structures, such as chiral charge-density waves~\cite{Lin104,Zenker2013,Zhang2024_frozenchiral}, or spin-density waves~\cite{yu2012}. This opens possibilities of using phonons to detect and manipulate generalized time-reversal symmetry breaking orders, enabling probes of otherwise hidden orders, as proposed in recent work~\cite{sutcliffe2025pseudo}.

Phonons exhibiting anisotropic gyromagnetic ratios in the simple and highly symmetric materials studied here suggest similar phenomena occur more broadly in noncentrosymmetric crystals. We expect these phonons to produce unique effects such as a planar Zeeman effect, in which the applied magnetic field is perpendicular to the angular momentum. {During the review of this manuscript, an orthogonal Einstein-de Haas effect was predicted for anisotropic gyromagnetic ratios of electrons \cite{Xue2026anisotropicgyromagneticratioorthogonal}. The anisotropic gyromagnetic ratios of phonons in our case could reciprocally produce an orthogonal phonon Barnett effect.} Together with divergent gyromagnetic ratios, these phonons raise fundamental question of whether angular momentum or the phonon magnetic moment is the key quantity in spin–phonon coupling. Decoupling these quantities are anticipated to have profound implications for ultrafast magnetism, spin transport~\cite{kim2023chiral, ohe2024chirality}, and phonon Hall effects \cite{zhang:2015,Grissonnanche2019,Grissonnanche2020,park2020phonon}. 

Anisotropic and divergent gyromagnetic ratios of phonons further point toward hidden orders and multipolar degrees of freedom encoded in lattice excitations~\cite{aeppli2020hidden}. For example, the phonons in Fig.~\ref{figGaAs} clearly lead to a magnetization with moments of higher order than the dipole, allowing them to couple to magnetic-field gradients and electric fields \cite{shitadeTheoryOrbitalMagnetic2018}. The interactions between axial phonons and spin-polarized electrons could be affected by such higher-order multipoles, especially in cases where the magnetic ions themselves exhibit multipolar order \cite{aeppli2020hidden, verbeek2023hidden}. Such magnetic multipoles associated with phonons could be detected experimentally in X-ray or neutron scattering experiments~\cite{urru2023neutron, soh2024spectroscopic}. 
 

\section*{Acknowledgments}
S.C. acknowledges support from  JSPS KAKENHI No. JP23H04865, MEXT, Japan. 
C.P.R. acknowledges support from the project FerrMion of the Ministry of Education, Youth and Sports, Czech Republic, co-funded by the European Union (CZ.02.01.01/00/22\_008/0004591), the European Union and Horizon 2020 through grant no.  810451, and ETH Zurich. D.M.J. acknowledges support from the ERC Starting Grant CHIRALPHONONICS, no. 101166037. Computational resources were provided by the Swiss National Supercomputing Center (CSCS) under project ID s1128.

\appendix


\section{Computational details}

The phonon energies and eigenvectors of monolayer h-BN were obtained using density functional perturbation theory with the Abinit software package (v. 10) \cite{gonze2020abinit}. Norm-conserving pseudopotentials were used as recieved from the Abinit library. The PBE GGA exchange--correlation functional \cite{perdew1996generalized} was used with the vdw-DFT-D3(BJ) dispersion correction of Grimme \cite{grimme2010consistent}. A plane-wave basis set energy cutoff of 30 Ha and a $16 \times 16\times 1$ grid of \textbf{k}-points were chosen following convergence studies. The structure was relaxed to an internal pressure of $-7$ MPa, and an $8 \times 8\times 1$ grid of \textbf{q}-points was then used for the phonon calculations. Computational data are publicly available from Ref. \cite{sdata}. DFT-calculated phonon energies and eigenvectors of GaAs were obtained from the Materials Project database \cite{jain2013commentary, osti_1200591}. The phonon angular momentum was obtained from the phonon eigenvectors following Eq.~\eqref{eq:phononangmom}; the phonon magnetic moments were obtained following Eq.~\eqref{eq:phononmagmom}.


\section{Group theoretical analysis of phonon modes in $\text{Co}_3\text{Sn}_2$S$_2$}
In this section, we present group theory based arguments for the existence of phonons modes with out-of-plane motion and finite PAM in Kagome Weyl semimetal  Co$_3$Sn$_2$S$_2$. This material crystallizes in the trigonal (rhombohedral) space group \(R\overline3m\) (No.\,166), adopting a shandite‑type structure. 
The Wyckoff positions in the hexagonal setting are Sn1 (site 3a), Sn2 (3b), S (6c), and Co (9d). Most interestingly,  Co atoms which reside on the 9d Wyckoff position are arranged on two-dimensional kagome lattices stacked along the crystal $c$ axis.  This structure gives rise to a total of 21 phonon modes with three acoustic and 18 optical modes which decompose into $A_{1g} \oplus E_g \oplus A_{1u} \oplus 5A_{2u} \oplus 6E_u$ irreducible representations.

We calculate the basis functions for $E_u$ modes using SAM on Bilbao crystallographic server~\cite{kroumova2003bilbao} and found that three $E_u$ mode basis functions involve the motion of Co ion as depicted in table~\ref{tab:Eutable}. The chiral superposition of $E_u(a)$
 and $E_u(b)$ modes result in in-plane circular motion for the first two basis vectors shown in the table. On the other hand, the third basis vector $E_u^3$ involves only out-of-plane motion. The chiral superposition of  $E_u^3(a)$
and $E_u^3(b)$ modes would result in a relative phase difference on Co atoms Co$_1$, Co$_2$, Co$_3$ located  at three positions  $(\tfrac{1}{2}, 0, \tfrac{1}{2})$, $(0, \tfrac{1}{2}, \tfrac{1}{2})$, and $(\tfrac{1}{2}, \tfrac{1}{2}, \tfrac{1}{2})$, respectively which are related by three-fold rotation about z-axis passing through $(0,0,0)$. It can be seen explicitly by writing the basis vectors $E_u^3$  in the basis of (Co$_1$, Co$_2$, Co$_3$):
\begin{equation}
    Q^{\pm}=E_u^3(a)\pm i E_u^3(b)=-\sqrt{\frac{2}{3}}\left(e^{-i2\pi/3}~\hat{z},~\hat{z},~e^{i2\pi/3}~\hat{z}\right)
\end{equation}
and under three-fold rotation $C_3(z)$ (Co$_1$, Co$_2$, Co$_3$)=(Co$_2$, Co$_3$, Co$_1$) which results in:
\begin{align}
    C_3(z)\left( Q^{\pm}\right)&=-e^{\pm i2\pi/3}\sqrt{\frac{2}{3}}\left(e^{-i2\pi/3}~\hat{z},~\hat{z},~e^{i2\pi/3}~\hat{z}\right)\nonumber\\
    &= e^{\pm i2\pi/3}Q^{pm}.
\end{align}
This shows how the out-of-plane basis vectors of $E_u$ modes in Co$_3$Sn$_2$S$_2$ can carry PAM.
 \begin{table}[h!]
\centering
\begin{tabular}{|c|c|c|c|c|c|c|c|}
\hline
Co atom & Position & $E_u^1(a)$ & $E_u^1(b)$ & $E_u^2(a)$ & $E_u^2(b)$ & $E_u^3(a)$ & $E_u^3(b)$ \\
\hline
$\mathbf{X}_1$ & $(\tfrac{1}{2}, 0, \tfrac{1}{2})$ & $\frac{1}{\sqrt{3}}$ & $\cdot$ & $\frac{1}{\sqrt{12}}$ & $\frac{\sqrt{3}}{\sqrt{12}}$ & $\cdot$ & $\cdot$\\
$\mathbf{Y}_1$ &  & $\cdot$ & $\frac{1}{\sqrt{3}}$ & $-\frac{\sqrt{3}}{\sqrt{12}}$ & $\frac{1}{\sqrt{12}}$ & $\cdot$ & $\cdot$\\
$\mathbf{Z}_1$ &  & $\cdot$ & $\cdot$ & $\cdot$ & $\cdot$ & $\frac{1}{\sqrt{6}}$ & $\frac{1}{\sqrt{2}}$\\
\hline
$\mathbf{X}_2$ & $(0, \tfrac{1}{2}, \tfrac{1}{2})$ & $\frac{1}{\sqrt{3}}$ & $\cdot$ & $\frac{-2}{\sqrt{12}}$  & $\cdot$ & $\cdot$ & $\cdot$\\
$\mathbf{Y}_2$ &  & $\cdot$ & $\frac{1}{\sqrt{3}}$ & $\cdot$ & $\frac{-2}{\sqrt{12}}$ & $\cdot$ & $\cdot$\\
$\mathbf{Z}_2$ &  & $\cdot$ & $\cdot$ & $\cdot$ & $\cdot$ & $\frac{-2}{\sqrt{6}}$  & $\cdot$\\
\hline
$\mathbf{X}_3$ & $(\tfrac{1}{2}, \tfrac{1}{2}, \tfrac{1}{2})$ & $\frac{1}{\sqrt{3}}$ & $\cdot$ & $\frac{1}{\sqrt{12}}$ & $-\frac{\sqrt{3}}{\sqrt{12}}$ & $\cdot$ & $\cdot$ \\
$\mathbf{Y}_3$ &  & $\cdot$ & $\frac{1}{\sqrt{3}}$ & $\frac{\sqrt{3}}{\sqrt{12}}$ & $\frac{1}{\sqrt{12}}$ & $\cdot$ & $\cdot$\\
$\mathbf{Z}_3$ &  & $\cdot$ & $\cdot$ & $\cdot$ & $\cdot$ &$\frac{1}{\sqrt{6}}$ &$-\frac{1}{\sqrt{2}}$\\
\hline
\end{tabular}
\caption{Basis functions for the $E_u$ irreducible representation at Wyckoff position 9d in $R\overline{3}m$.}
\label{tab:Eutable}
\end{table}


\section{General derivation of effective phonon magnetic moments from orbit-lattice coupling}

In this section, we follow Ref.~\cite{chaudhary2024giant} to calculate the splitting of axial phonons.  We begin by considering a degenerate phonon mode with two components that is described by the Hamiltonian
    \begin{equation}
        H_{ph}=\omega_0(a^\dagger a+ b^\dagger b).
    \end{equation}
We only consider phonon modes near the Brillouin-zone center and can accordingly drop the momentum dependence in the phonon operators and energies. In order to account for the effect of orbital-lattice coupling on the phonon spectrum, we use a Green's function formalism. For the non-interacting system, the Green's function matrix is given by
    
    \begin{equation}
        \mathbf{D}_0(\omega)=\begin{pmatrix}
        D_0^{aa}(\omega)&0\\0&D_0^{bb}(\omega)
        \end{pmatrix},
    \end{equation}
    where the components are given by
    \begin{equation}
        D_0^{aa}(\omega)=D_0^{bb}(\omega)=\frac{2\omega_0}{\omega^2-\omega_0^2}.
    \end{equation}
The phonon frequency, $\omega_0$, can be trivially retrieved by solving $\mathrm{Det}(\mathbf{D}_0^{-1}(\omega))=0$.

We next consider the electronic Hamiltonian
    \begin{equation}
        H_{el}=\sum_{i=1}^4\varepsilon_ic_i^\dagger c_i,
    \end{equation}
    where $c_i^\dagger$  and $c_i$ are the creation and annihilation operators for electrons in state $i$ on the magnetic ion. We focus on the case where these states are  two Kramers doublets represented by states $(\ket{1},\ket{2})$ and $(\ket{3},\ket{4})$.
    
Next, we consider an orbit-lattice interaction of this form:
\begin{equation}
    H_{el-ph} =V^a+V^b,
\end{equation}
where
\begin{align}
        V^a=\sum_{i,j}g(a^\dagger+a)\Gamma^a_{ij}&=g(a^\dagger+a)(c_3^\dagger c_1+c_1^\dagger c_3)\nonumber\\
        & -g(a^\dagger+a)(c_4^\dagger c_2+c_2^\dagger c_3), \label{vertices1}\\
        V^b=\sum_{i,j}g(a^\dagger+a)\Gamma^a_{ij}&=ig(b^\dagger+b)(c_3^\dagger c_1-c_1^\dagger c_3)\nonumber\\
        &+ig(b^\dagger+b)(c_4^\dagger c_2-c_4^\dagger c_2).\label{vertices2}        
\end{align}

The new phonon Green's function after including these interactions is:
\begin{widetext}
\begin{equation}
    \mathbf{D}^{-1}=\begin{pmatrix}
    \frac{\omega^2-\omega_0^2}{2\omega_0}-\tilde{g}^2\left(\frac{f_1\Delta_1}{\omega^2-\Delta_1^2}+\frac{f_2\Delta_2}{\omega^2-\Delta_2^2}\right)&i\tilde{g}^2\left(-\frac{f_1\omega}{\omega^2-\Delta_1^2}+\frac{f_2\omega}{\omega^2-\Delta_2^2}\right)\\
    -i\tilde{g}^2\left(-\frac{f_1\omega}{\omega^2-\Delta_1^2}+\frac{f_2\omega}{\omega^2-\Delta_2^2}\right)&\frac{\omega^2-\omega_0^2}{2\omega_0}-\tilde{g}^2\left(\frac{f_1\Delta_1}{\omega^2-\Delta_1^2}+\frac{f_2\Delta_2}{\omega^2-\Delta_2^2}\right)
    \end{pmatrix},
    \label{Dvals}
\end{equation}
\end{widetext}
where $\tilde{g}^2=2 g^2$, $\Delta_1=\varepsilon_{31}$, $\Delta_2=\varepsilon_{42}$, $f_i$ is the occupation of $i^{th}$electronic  band and we assume the excited state to be unoccupied, $f_3=f_4=0$. The modified energies can then be obtained by solving $\mathrm{Det}(\mathbf{D}^{-1})=0$.

When a magnetic field $\mathbf{B}=B\,\hat{z}$ is applied, electronic transition energies are modified as follows:
\begin{equation}
    \Delta_1=\Delta-\gamma B,~~\Delta_2=\Delta+\gamma B,
\end{equation}
where $\gamma=\mu^{el}_{ex}-\mu^{el}_{gs}$ depends on the magnetic moment of the ground- and excited-state doublets. Lifting the degeneracy of the ground-state doublet leads to  asymmetric populations of the ground-state energy levels, $f_{12}\neq 0$. 
 Accordingly, the secular equation, $\mathrm{Det}(\mathbf{D}^{-1}(\omega))=0$ for $\mathbf{D}^{-1}$ given by Eq.~\eqref{Dvals}  becomes:
\begin{widetext}
\begin{align}
    &(\omega^2-\omega_0^2)(\omega^2-\Delta^2)-2\tilde{g}^2f_0\omega_0\Delta+
    2 \omega \left(B\gamma (\omega^2-\omega^2_0)+\tilde{g}^2\omega_0f_{21}\right)
    +\gamma B\left(\gamma B (\omega^2-\omega^2_0)+2\tilde{g}^2\omega_0f_{21}\right)=0,\label{sc1}\\
    &(\omega^2-\omega_0^2)(\omega^2-\Delta^2)-2\tilde{g}^2f_0\omega_0\Delta
    -2 \omega \left(B\gamma (\omega^2-\omega^2_0)+\tilde{g}^2\omega_0f_{21}\right)
    +\gamma B\left(\gamma B (\omega^2-\omega^2_0)+2\tilde{g}^2\omega_0f_{21}\right)=0.
    \label{seculareq}
\end{align}
\end{widetext}
These two equations are not equivalent and there is a term linear in $\omega$ that indicates a frequency splitting of phonon and electronic excitations. Given that the orbit-lattice coupling is weak and the electronic excitations are off-resonant from phonons, we can assume that phonon energies are modified only slightly and have the form $  w_{ph}^{\pm}=\Omega_{ph}\left(1\mp \eta \right)$ which gives:
\begin{align}
    \Omega_{ph}\eta = & \frac{\gamma B(\Omega_+^2-\omega^2_0)+\tilde{g}^2\omega_0 f_{21}}{\Omega_{ph}^2-\Omega_{{el}}^2+\gamma^2B^2} \nonumber\\
    = & \frac{\gamma B(\Omega_+^2-\omega^2_0)+\tilde{g}^2\omega_0\tanh\left(\frac{\mu^{el}_\mathrm{gs}B}{k_BT}\right)}{\sqrt{\left(\omega_0^2-\Delta^2\right)^2+8\tilde{g}^2f_0\omega_0\Delta}+\gamma^2B^2}.
\end{align}
where $f_{21}$was replaced by $ -\tanh\left(\frac{\mu^{el}_{gs} B}{k_BT}\right)$  with $\mu^{el}_{gs}$ representing the magnetic moment of ground state manifold as the system is paramagnetic. For the off-resonant case, we can assume $|\Delta-\omega_0|\gg \gamma B$ and therefore neglect the linear $B$ term in the numerator and the quadratic one in the denominator. The off-resonant case is a reasonable assumption, as $\gamma B\sim 0.5$~meV in strong magnetic fields of $B=10$~T, whereas often $|\Delta-\omega_0|>10$~meV. As a result, the splitting of the phonon frequencies can be written as
\begin{equation}
   \frac{\omega_{ph}^+-\omega_{ph}^-}{\omega_{ph}(B=0)}\approx\frac{2\tilde{g}^2}{\sqrt{\left(\omega_0^2-\Delta^2\right)^2+8\tilde{g}^2f_0\omega_0\Delta}}\tanh\left(\frac{\mu^{el}_{gs}B}{k_BT}\right),
   \label{split1}
\end{equation}
which is the main expression used to evaluate the splitting shown in Fig.~\ref{figCeCl3Eg}.


\section{Symmetry properties and $E_{2u}$ modes of $\text{CeCl}_3$}
\begin{figure*}
    \centering
    \includegraphics[scale=0.45]{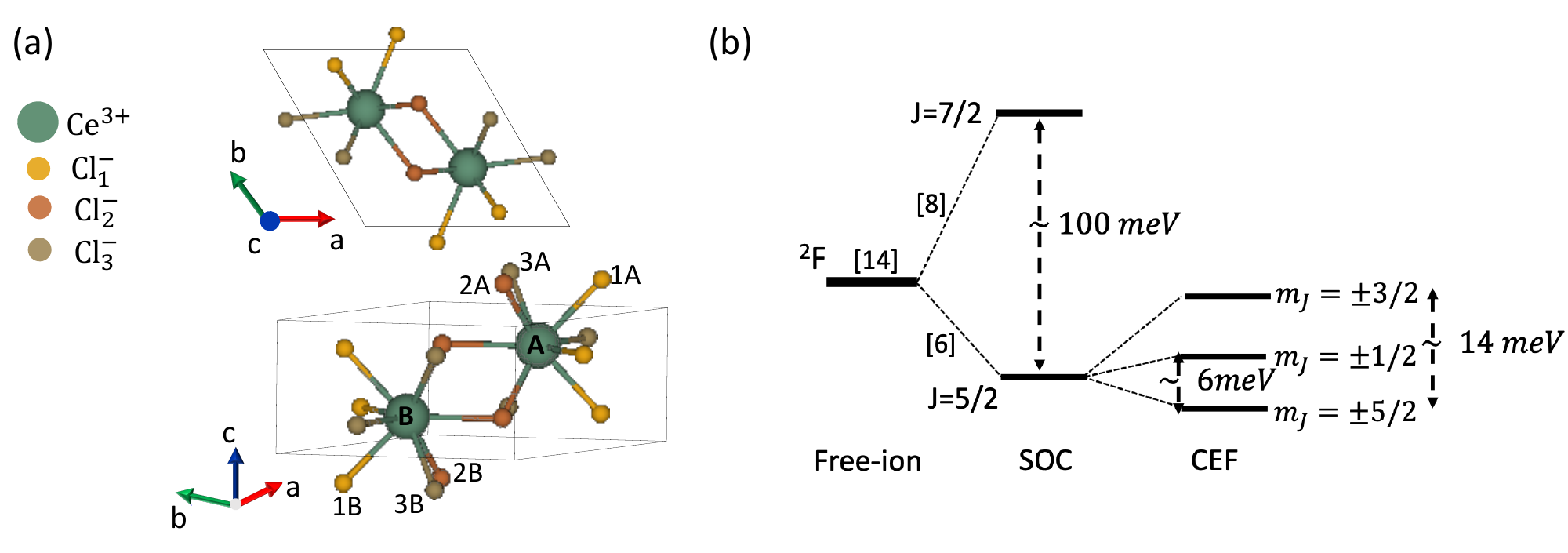}
    \caption{(a) CeCl$_3$ crystal structure along two directions. (b) Electronic energy levels of the Ce$^{3+}$ ion. The energy levels compared to the free ion are split by spin-orbit coupling and by the crystal electric field, resulting in three Kramers doublets, of which $\pm5/2$ is the ground state. }
    \label{figCeCl3}
\end{figure*}

The rare-earth trihalide CeCl$_3$, shown in  Fig.~\ref{figCeCl3}(a), crystallizes in the space group no. 176 (point group $6/m$) and its primitive unit cell contains eight atoms.  The two Ce$^{3+}$ ions are located at the 2$c$ Wyckoff positions and the six Cl$^{-}$ ions are located at the 6$h$ Wyckoff positions. 
Each Ce$^{3+}$ ion has nine nearest neighbors arranged in three different planes. In each plane, the three Cl$^{-}$ atoms are related by a three-fold rotation along a $z$ axis passing through Ce$^{3+}$ ion or the center of the hexagon.
This structure leads to 21 optical phonon modes out of which two modes ( $E_{1g}$ and $E_{2u}$) have purely out-of-plane motion ~\cite{bcs:sam} consisting of the irreducible representations $2A_g\oplus 1A_u 2B_g\oplus2B_u\oplus1E_{1g}\oplus3E_{2g}\oplus2E_{1u}\oplus1E_{2u}$~\cite{Juraschek2022_giantphonomag}. Using SAM on Bilbao crystallographic server~\cite{bcs:sam}, we found that out of these seven doubly degenerate phonons , two modes ( $E_{1g}$ and $E_{2u}$) have purely out-of-plane motion.

The ground-state configuration of the Ce$^{3+}$ ($4f^1$) ion is given by a nearly free-ion configuration of a $L=3,S=1/2$ state in accordance with Hund's rule. The spin-orbit coupling splits this 14 dimensional space into $J=5/2$ and $J=7/2$ total angular momentum sectors and the ground-state is given by the six-dimensional $J=5/2$ $(^2F_{5/2})$ state as shown in Fig.~\ref{figCeCl3} (b). Since there is only one electron in the 4$f$ orbitals, the wavefunctions of different states in this multiplet can be written as
    \begin{align}
    \ket{J=5/2,m_j=\pm5/2}=&-\sqrt{\frac{1}{7}}\ket{m_l=\pm2,m_s=\pm1/2}\nonumber\\
    &+\sqrt{\frac{6}{7}}\ket{m_l=\pm3,m_s=\mp1/2},\label{state52}\\
    \ket{J=5/2,m_j=\pm3/2}=&-\sqrt{\frac{2}{7}}\ket{m_l=\pm1,m_s=\pm1/2}\nonumber\\
    &+\sqrt{\frac{5}{7}}\ket{m_l=\pm2,m_s=\mp1/2},\label{state32}\\
    \ket{J=5/2,m_j=\pm1/2}=&-\sqrt{\frac{3}{7}}\ket{m_l=\pm0,m_s=\pm1/2}\nonumber\\
    &+\sqrt{\frac{4}{7}}\ket{m_l=\pm1,m_s=\mp1/2},\label{state12}
    \end{align}
where $\ket{m_l,m_s}$ is a 4$f$ orbital state with orbital quantum number $m_l$ and spin quantum number $m_s$. The CEF further splits the states into three Kramers doublets $\ket{\pm5/2}$, $\ket{\pm1/2}$, and $\ket{\pm3/2}$ with energies 0~meV, 5.82~meV, 14.38~meV, respectively~\cite{schaack:1977}.

We use a point-charge model to describe the crystal electric field of the system, in which the potential energy of an electron at position $\mathbf{r}$ from Ce$^{3+}$ nucleus due to the $n$-th Cl$^-$ ion is given by
\begin{equation}
    V(\mathbf{R}_n, \mathbf{r})=\frac{e^2}{4\pi\epsilon_0}\frac{1}{|\mathbf{R}_n-\mathbf{r}|},
    \label{Coulomb}
\end{equation}
where $\mathbf{R}_n=\mathbf{R}_{0,n}+\mathbf{u}_{n}$ is the displacement of the $n$-th ligand ion from Ce$^{3+}$ nucleus, which depends on the equilibrium displacement, $\mathbf{R}_{0,n}$, and the relative lattice displacement, $\mathbf{u}_{n}$, arising from the phonon.

The perturbation introduced by a given phonon mode can be obtained by a Taylor expansion of the potential in the lattice displacement $\mathbf{u}_{n}$ to linear order, which is done in Mathematica using the built-in series expansion function which evaluates 
\begin{equation}
\partial_{u_n^\alpha}\partial_{r^\beta}\partial_{r^\gamma}   \left(\frac{1}{|\mathbf{R}_n-\mathbf{r}|}\right)\bigg |_{\mathbf{R}_n=\mathbf{R}_{0,n},\mathbf{r}=0}.
\end{equation}

The $E_{2u}$ phonon lowers the symmetry around the magnetic ion and for the lattice distortion induced by this phonon, the first order term for the change in Coulomb potential is given by: 
\begin{align}
V(E_{1g}(a))&=\left[-0.1 xz+0.066 yz\right]Q_a~\frac{\mathrm{eV}}{\text{\AA}^3\sqrt{\mathrm{amu}}}, \\
    V(E_{1g}(b))&=\left[0.066 xz+0.1 yz\right]Q_b~\frac{\mathrm{eV}}{\text{\AA}^3\sqrt{\mathrm{amu}}}.
    \label{Eq:coupling}
\end{align} 
Now, we can express $xz=r^2\sin\theta\cos\theta \cos\phi$ and $yz=r^2\sin\theta\cos\theta \cos\phi$ in spherical coordinates. The electronic states on Ce$^{3+}$ ion can be written in terms of $\ket{L=3,m=m_l}$ which have wavefunction $\langle r|L=3,m=m_l \rangle =R (r) Y_3^{m_l}(\theta,\phi)$. This allows us to calculate the matrix elements between different 4$f$ states and the only nonzero terms are given by
\begin{align}
\langle m=\pm 3|xz|m=\pm 2\rangle=\mp \langle r^2\rangle\frac{1}{3\sqrt{6}},\\
\langle m=\pm 2|xz|m=\pm 1\rangle=\mp \langle r^2\rangle\frac{1}{3\sqrt{10}},\\
\langle m=\pm 1|xz|m=\pm 0\rangle=\mp \langle r^2\rangle\frac{1}{3\sqrt{75}},
\end{align}

\begin{align}
\langle  m=\pm3|yz|m= \pm 2\rangle= \langle r^2\rangle\frac{i}{3\sqrt{6}},\\
\langle  m=\pm2|yz|m=\pm 1\rangle=\langle r^2\rangle\frac{i}{3\sqrt{10}},\\
\langle m= \pm 1|yz|m=0\rangle= \langle r^2\rangle\frac{i}{3\sqrt{75}}.
\end{align}
Using these values for states given in Eq.~\eqref{state52}-\eqref{state12}, we obtain the matrices for $xz$ and $yz$ operators
\begin{align}
    H_1(xz)&=-\frac{2}{7\sqrt{5}}\langle r^2\rangle\begin{pmatrix}
    &\left|\frac{5}{2},\pm\frac{5}{2}\right>&\left|\frac{5}{2},\pm\frac{3}{2}\right>\\
    \left|\frac{5}{2}\pm\frac{5}{2}\right>&0&\pm 1\\[1.2em]\left|\frac{5}{2},\pm\frac{3}{2}\right>&\pm 1&0
    \end{pmatrix},
    \\
     H_1(yz)&=\frac{2}{7\sqrt{5}}\langle r^2\rangle\begin{pmatrix}
    &\left|\frac{5}{2},\pm\frac{5}{2}\right>&\left|\frac{5}{2},\pm\frac{3}{2}\right>\\
    \left|\frac{5}{2},\pm\frac{5}{2}\right>&0&i\\[1.2em]\left|\frac{5}{2},\pm\frac{3}{2}\right>&-i&0
    \end{pmatrix}
    \label{Eq:matrix}
\end{align}
where $\langle r^2\rangle=\int_0^\infty r^2|R(r)|^2r^2dr$ is the mean-square 4$f$-electron radius. We can write the phonon displacements as
\begin{align}
    Q_a  &=\frac{\hbar}{\sqrt{\hbar\omega_{0}}}\left(a+a^\dagger\right)=\frac{0.06\text{\AA}\sqrt{\mathrm{eV\,amu}}}{\sqrt{\hbar\omega_{{0}}}}\left(a+a^\dagger\right), \\
    Q_b  &=\frac{\hbar}{\sqrt{\hbar\omega_{0}}}\left(b+b^\dagger\right)=\frac{0.06\text{\AA}\sqrt{\mathrm{eV\,amu}}}{\sqrt{\hbar\omega_{{0}}}}\left(b+b^\dagger\right), \label{Qba}
\end{align}
where we restored $\hbar$ and $\hbar\omega_{{0}}$ is the energy of phonon mode. As a result, the orbit-lattice coupling operators now takes the following form

\begin{equation}
H_{el-ph}=(a^\dagger+a)\hat{O}_a+(b^\dagger+b)\hat{O}_b,
\label{elph1}
\end{equation}
where
\begin{align}
\hat{O}_a & =ge^{i\theta}\bigg|+\frac{5}{2}\bigg\rangle\bigg\langle+\frac{3}{2}\bigg|-ge^{-i\theta}\bigg|-\frac{5}{2}\bigg\rangle\bigg\langle-\frac{3}{2}\bigg|+\mathrm{h.c.},\label{couplingE1ga}\\
\hat{O}_b & =ige^{i\theta}\bigg|+\frac{5}{2}\bigg\rangle\bigg\langle+\frac{3}{2}\bigg|+ige^{-i\theta}\bigg|-\frac{5}{2}\bigg\rangle\bigg\langle-\frac{3}{2}\bigg|+\mathrm{h.c.}
\label{couplingE1gb}
\end{align}
Here, we combined Eqs.~(\ref{Eq:coupling}), (\ref{Eq:matrix}), and (\ref{Qba}) in order to obtain $g = -\sqrt{0.1^2+0.066^2}\frac{2}{7\sqrt{5}}\langle r^2\rangle \frac{0.06}{\sqrt{\omega_0}} {\mathrm{eV}}^{3/2}/\text{\AA}^2$ 
and $\tan(\theta)=0.66/0.1$. This analysis shows that $E_{2u}$ mode ($\omega_0$=20.5 meV) couples  with CEF excitation between $\ket{\pm3/2}$ and $\ket{\pm5/2}$ which has energy $\Delta= 14$ meV. Using the value of $g$, we can now evaluate the splitting of two axial phonons using Eq.~\eqref{split1} and with $\mu_{gs}^{el}=2\mu_B$ which is the magnetic moment for $J=5/2, m_j=\pm 5/2$ which comes out to be
\begin{equation}
   \omega_{ph}^+-\omega_{ph}^-\approx 0.3\tanh\left(\frac{2\mu_BB}{k_BT}\right) \,\,\text{meV}
   \label{splitting_value}
\end{equation}
which indicates a saturation splitting close to 0.3 meV as shown in Fig.~\ref{figCeCl3Eg}(c).


\bibliography{ref}

\end{document}